\documentclass[prb,twocolumn,showpacs,preprintnumbers,amsmath,amssymb]{revtex4}
\usepackage{graphicx,verbatim}
\usepackage{color,subfigure}

\def\bcalG{\boldsymbol{\mathcal{G}}}
\def\brho{\boldsymbol{\rho}}
\def\bgamma{\boldsymbol{\gamma}}
\def\stk#1#2{{\begin{smallmatrix}\\#1\\#2\end{smallmatrix}}}

\newcommand{\xx}{\mathbf{x}}
\newcommand{\kk}{\mathbf{k}}
\newcommand{\pp}{\mathbf{p}}
\newcommand{\qq}{\mathbf{q}}
\newcommand{\QQ}{\mathbf{Q}}
\newcommand{\0}{\mathbf{0}}
\newcommand{\vare}{\varepsilon}

\newcommand{\tvare}{\widetilde{\vare}}
\newcommand{\tDelta}{\widetilde{\Delta}}
\newcommand{\calD}{\mathcal{D}}

\newcommand{\calV}{\mathcal{V}}

\newcommand{\half}{\frac{1}{2}}
\newcommand{\etal}{{\em et al.}}

\begin{document}
\title{Effect of Spin Fluctuations on Phonon-Mediated Superconductivity in the Vicinity of a Quantum Critical Point}
\author{Y. L. Loh}
\affiliation{Department of Physics, Purdue University, West Lafayette, IN  47907 }
\author{P. B. Littlewood}
\affiliation{Theory of Condensed Matter Group, Cavendish Laboratory, University of Cambridge, Madingley Road, Cambridge CB3 0HE, United Kingdom}
\date{\today}
\begin{abstract}
We consider an s-wave superconductor in the vicinity of a second-order ferromagnetic (FM) or spin-density-wave (SDW) quantum critical point (QCP), where the superconductivity and magnetism arise from separate mechanisms.  The quantum critical spin fluctuations reduce the superconducting $T_c$.  Near a FM QCP, we find that $T_c$ falls to zero as $1/\ln (1/\kappa)$ in 3D and as $\kappa$ in 2D, where $\kappa \sim |J-J_c|^\nu$ is the inverse correlation length of the spin fluctuations, and measures the distance $|J-J_c|$ from the quantum critical point.  SDW quantum critical fluctuations, on the other hand, suppress $T_c$ to zero as $\sqrt{\kappa}$ in 2D, and suppress $T_c$ only to a finite value in 3D, producing a cusp of the form $\text{const} + |J-J_c|^\nu$.
\end{abstract}
\maketitle

\section{Introduction}
There has been much interest recently in the interplay of superconductivity and magnetic phenomena.  Many different scenarios arise depending on the superconducting pairing symmetry (singlet/triplet, $s$/$p$/$d$-wave)
\footnote{Strictly speaking, the angular momentum classifications $s$,$p$,$d$ should be replaced by the appropriate representations of the symmetry group of the crystal.}
 and the type of magnetism (static/fluctuating ferromagnetism/antiferromagnetism/spin density waves).  Much  attention has been given to magnetically-mediated unconventional superconductivity, such as models of d-wave singlet superconductivity driven by antiferromagnetic spin fluctuations (possibly applicable to the cuprates and heavy-fermion superconductors\cite{Miyake86}), and $p$-wave-like triplet pairing in He$^3$ \cite{anderson1961,balian1963}, Sr$_2$RuO$_4$ \cite{maeno1994} and UGe$_2$ \cite{Saxena00,Saxena01} driven by ferromagnetic spin fluctuations.  

However, there are of course systems where conventional $s$-wave pairing coexists with magnetism.  While the coexistence of antiferromagnetic or SDW order with superconductivity is not uncommon, the coexistence of FM with $s$-wave superconductivity is rare.  In ErRh$_4$B$_4$\cite{Fertig77} and HoMo$_6$S$_8$\cite{Lynn81,Ishikawa77}, magnetism and superconductivity arise from independent mechanisms (RKKY coupling between $f$-electrons and phonon exchange respectively).  It is now well accepted that \emph{static magnetic order} has a pair-breaking effect on BCS singlet $s$-wave superconductivity (as does any perturbation which breaks time-reversal symmetry), often resulting in the complete destruction of superconductivity.  The mean-field theory of ferromagnetic superconductors was studied by Gor'kov and Rusinov \cite{Gorkov64}, although a more general analysis suggests a Larkin-Ovchinnikov-Fulde-Ferrell state with a non-uniform order parameter \cite{Larkin65,Fulde64}.  A mean-field theory based on a simple `nesting-fraction' model (similar to that of Bilbro and McMillan \cite{Bilbro76} for the coexistence of charge-density-wave order and superconductivity) shows that coexistence of spin-density-wave (SDW) order and superconducting order may also be possible in a limited range of parameters\cite{Saravanamuttu}. 

One expects that magnetic \emph{fluctuations} should also have a pair-breaking effect on $s$-wave superconductivity, especially for a system close to a second-order magnetic QCP, where the amplitude of the fluctuations becomes very large.  Berk and Schrieffer\cite{Berk66} showed numerically that in the presence of ferromagnetic fluctuations a finite electron-phonon coupling $V_\text{ph}$ is required to produce superconductivity, and in Pd, for example, this critical coupling is so large that superconductivity does not occur at all.  We study this problem in more detail, obtaining analytic results for the dependence of $T_c$ as the QCP is approached.  

We assume that the pairing interaction is due to phonon exchange, and can be represented by a BCS interaction, whereas the magnetism is driven by a coupling $J$ between itinerant electrons or localized spins such that when $J$ exceeds a critical coupling $J_c$ at $T=0$ the system undergoes a second-order phase transition from a disordered state to a magnetically ordered state.  The resulting interaction between electron spins (as a function of Matsubara frequency $i\omega_n$) is assumed to be of the `Hertzian' form 
	\begin{align}
 	\calV^s _{n\qq}
	&\approx
		\frac{J {\kappa_0}^2}{\kappa^2 + q^2 + \frac{|\omega_n|}{\alpha q}} 
	~
	\Bigg(
	\text{or~}
		\frac{J {\kappa_0}^2}{\kappa^2 + |\qq-\QQ|^2 + \frac{|\omega_n|}{\alpha}}
	\Bigg)
	\end{align}
near a FM(SDW) QCP.  Here $\kappa^{-1}$ is the diverging correlation length.
This approach is similar to the fluctuation-exchange (FLEX) approach used by Monthoux \cite{Monthoux92} to study spin-fluctuation-mediated $d$-wave-like superconductivity, and to that of Wang \etal\cite{Wang01} and Roussev and Millis\cite{Roussev01}, who studied a model of $p$-wave pairing where the ferromagnetic fluctuations suppressed $T_c$ to a \emph{finite value} at the QCP.  Also, Li \etal \cite{li2006} have performed related studies of the electromagnetic properties of superconductors near ferromagnetic instabilities.

As the QCP is approached, the energy scale of the spin fluctuations goes to zero (`critical slowing down') and their amplitude goes up.    We show that in the vicinity of the quantum critical point, pair-breaking by slow spin fluctuations can be mapped onto the Abrikosov-Gor'kov problem of pair-breaking by static magnetic impurities.  We thus predict that $T_c$ may or may not be suppressed all the way to zero at the QCP, depending upon the dimensionality and the type of magnetism:
\begin{center}
\begin{tabular}{lll}
2D FM  &~~~& $T_c \sim \kappa           \sim |J-J_c|^\nu$ \\
2D SDW && $T_c \sim \sqrt{\kappa}    \sim |J-J_c|^{\nu/2}$ \\
3D FM  && $T_c \sim 1/\ln(1/\kappa)        \sim 1/\ln (1/|J-J_c|)$ \\
3D SDW && $T_c \sim \text{const} +  |J-J_c|^\nu$ \\
\end{tabular}
\end{center}
\noindent 
where $\nu$ is the correlation length exponent.
We expect these forms to apply on both sides of the transition.

\section{Charge-charge interaction}
Starting with an electron-phonon Hamiltonian, integrating out the phonons in the harmonic approximation, and making certain approximations for the structure of the electron-phonon coupling in momentum space leads to the following action for phonon-mediated superconductivity:
	\begin{align}
	S[\psi,\bar\psi]
	&=T\sum_{n\alpha} \int_\kk (i\vare_n - \xi_\kk) 
		\bar\psi_{n\kk\alpha} \psi_{n\kk\alpha}
		\nonumber\\&
	+
		\int_{\tau\tau'\xx\xx'}
		\tfrac{1}{2}
		\calV^c_{\tau-\tau',\xx-\xx'}
		\rho_{\tau\xx} \rho_{\tau'\xx'}
	\label{e:action_kinetic_charge}
	\end{align}
where $\vare_n=2\pi(n+\half)T$ are fermionic Matsubara frequencies, $\alpha=\pm\half$ is a spin index, $\xi_\kk$ is the electron dispersion relation, $\tau$ runs from $0$ to $\beta=1/T$, $\rho_{\tau\xx} \equiv \sum_\alpha \tfrac{1}{2} \bar\psi_{\tau\xx\alpha} \psi_{\tau\xx\alpha}$ is the electron density operator, and $\calV^c$ is an effective interaction in the `charge' channel that is second order in the electron-phonon coupling $g$ and first order in the phonon propagator $\calD$.  If the superconductivity is isotropic and the Fermi surface is spherical, or if the system is dirty enough that the order parameter $\Delta$ is effectively averaged over the Fermi surface, then for the purpose of calculating superconducting properties $\calV^c_{nq}$ (the Fourier transform of $\calV^c_{\tau\xx}$) can be replaced by an effective interaction $\calV^c_n$ in the Eliashberg equations.  $\calV^c_n$ is usually referred to in the literature as ``$\alpha^2F$'', and is obtained by averaging the interaction $\calV^c_{nq}$ over pairs of momenta on the Fermi surface\cite{McMillan64}:
	\begin{align}
 	\calV^c_n
	&=
		\int_{\kk\pp}
		\delta(\xi_\kk)~
		\delta(\xi_\pp)~
		\calV^c_{n\kk\pp}
		\Bigg/
		\int_\kk
		\delta(\xi_\kk)~		
	\label{e26}
	\end{align}
where $\int_\kk \equiv \int \frac{d^d\kk}{(2\pi)^d}$ is an integral over the Brillouin zone.  Note that $\int_\kk \delta(\xi_\kk) \equiv \int_S \frac{d^{d-1}k}{(2\pi)^d v_F}$, where the integral is performed over the Fermi surface, weighted by the inverse Fermi velocity $1/v_F(\kk)$, where $v_F(\kk)=\left| \frac{\partial \xi(\kk)}{\partial \kk} \right|$.

\section{Effect of FM fluctuations on $T_c$}
\subsection{Spin-spin interaction $\calV^s _{n\qq}$ near a FM-QCP}
In the random phase approximation (RPA), the spin correlation function in an itinerant ferromagnet has the following form at small $n$ and $q$:
	\begin{align}
 	\chi _{nq}
	&=
	\frac{\chi^0 _{00} {\kappa_0}^2}{\kappa^2 + q^2 + \frac{|\omega_n|}{\alpha q}}
	\label{e:hertzian_chi}
	\end{align}
where $\kappa$, the inverse spin-fluctuation correlation length in the interacting system, is reduced from its bare value $\kappa_0$ as $\kappa = {\kappa_0} \left( 1 - J/J_c  \right)^\nu = 1/\xi$ with $\nu=1/2$, so that $\chi$ undergoes Stoner enhancement.   $\kappa$ indicates the `distance' from the ferromagnetic quantum critical point (QCP); a second-order quantum phase transition occurs at $J=J_c$, when $\kappa=0$.
Eq.~\eqref{e:hertzian_chi} can also be obtained by integrating out the electrons to obtain a Landau-Ginzburg-Wilson effective action for a spin field.

For a long time it was thought that low-energy long-wavelength quantum fluctuations are irrelevant (in $d>1$) so that Eq.~\eqref{e:hertzian_chi} is universally true\cite{Hertz76}. However, there is growing evidence that these `soft modes' are actually relevant, changing the form of $\chi$ or even causing the second-order QCP to become weakly first order \cite{Belitz99,belitz2005,rech2006}.
In this paper we restrict ourselves to regimes in which such effects is negligible, so that Eq.~\eqref{e:hertzian_chi} holds, albeit for some $\nu$ not necessarily equal to $1/2$.


This form for $\chi$ leads to an effective spin-spin \emph{interaction}	\begin{align}
 	\calV^s _{nq}
	&\approx
		\frac{J {\kappa_0}^2}{\kappa^2 + q^2 + \frac{|\omega_n|}{\alpha q}}
	\label{e:VsnqFM}
	\end{align}
which diverges at small $\omega_n$ and $q$ as $J\rightarrow J_c$; these large spin fluctuations are likely to have an important effect on superconductivity.

Eq.~\eqref{e:VsnqFM} is a generic expression for the spin-spin interaction near any kind of ferromagnetic QCP, and should be applicable regardless of whether the ferromagnetism and superconductivity arise from a single band of itinerant electrons or from different bands (or even if the magnetism is due to localized spins, provided that one can find a model of localized spins that exhibits a second-order quantum phase transition).
\footnote{The Stoner model described above exhibits a quantum critical point and a vertical FM phase boundary.  The standard Heisenberg model, however, gives a finite-temperature phase transition whose Curie temperature is proportional to $J$.}
Ultimately, the spin-fluctuation-mediated interaction can be represented by a low-energy effective action of a form analogous to Eq.~\eqref{e:action_kinetic_charge}:
	\begin{align}
	S^\text{spin}_\text{eff} [\psi,\bar\psi]
	&=
	\int_{\tau\tau'\xx\xx'}
	\tfrac{1}{2} \calV^s_{\tau-\tau', \xx-\xx'}
	\mathbf{S}_{\tau'\xx'} \cdot \mathbf{S}_{\tau\xx}
	.
	\label{e:action_spin}
	\end{align}

\subsection{Jacobian $\eta_\qq$ for spherical Fermi surface}
The dimensionless effective spin-spin interaction, $\calV^s_n$, may be calculated in the same way as the effective phonon-mediated attraction in Eq.~\eqref{e26}:
	\begin{align}
 	\calV^s_n
	&=
		\int_{\kk\pp}
		\delta(\xi_\kk)~
		\delta(\xi_\pp)~
		\calV^s_{n,\kk-\pp}
		\Bigg/
		\int_\kk
		\delta(\xi_\kk)	
	=
		\int_\qq
		\eta_\qq~ \calV^s_{n\qq}
	\label{e18}
	\end{align}
where $\eta_\qq$ is a Jacobian corresponding to the change of variables, which is the `autocorrelation of the Fermi surface':
	\begin{align}
 	\eta_\qq
	&=\int_\kk \delta(\xi_\kk)~ \delta(\xi_{\kk+\qq})
		\bigg/ \int_\kk \delta(\xi_\kk)  .
	\label{e37}
	\end{align}

	\begin{figure}
	\includegraphics{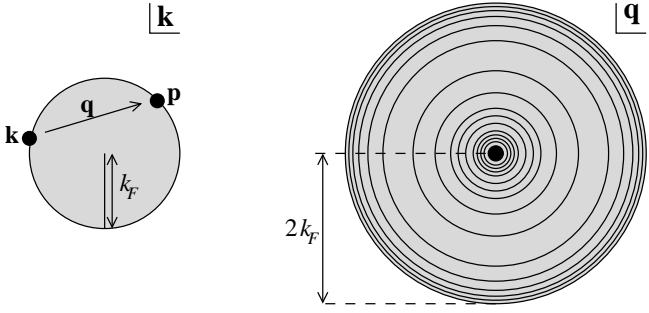}
	\caption{(Left) 2D circular Fermi surface in $\kk$-space. (Right) contours of $\eta(\qq)$ (schematic).}
	\end{figure}

$\eta$ can be calculated geometrically in 2D (3D) by considering the areas (volumes) of intersection of circular annuli (spherical shells).  The results are
	\begin{align}
 	\eta_\qq	
	&=\frac{m~ \Theta(1 - \frac{q}{2k_F}) }
			{\pi q k_F \sqrt{1 -  \big( \frac{q}{2k_F} \big)   ^2}}
		&\text{(2D)},
	\label{e:eta_2d}
\\
 	\eta_\qq	
	&=\frac{m~ \Theta(1 - \frac{q}{2k_F}) }{2qk_F}
		&\text{(3D)}.
	\label{e:eta_3d}
	\end{align}
Both these expressions are normalized such that $\int_\qq \eta_\qq = \nu$.
In both 2D and 3D, it can be shown that $\eta_\qq$ goes as $1/q$ near $\qq=\mathbf{0}$.  In both 2D and 3D this, by itself, is an integrable singularity.  However, when $\kappa=0$ and $\omega_n=0$, $V^s_{nq}$ goes as $1/q^2$ near $\qq=\mathbf{0}$.  The entire integrand of Eq.~\eqref{e18} thus goes as  $1/q^3$.  We shall show that this makes the integral divergent.

\subsection{Effective interaction $\calV^s_n$ near FM-QCP}
For the case of a 2D circular Fermi surface, substituting Eq.~\eqref{e:eta_2d} into \eqref{e18} gives
	\begin{align}
 	\calV^s_n
	&\approx
		\frac{mJ {\kappa_0}^2}{2\pi^2 k_F}
		\int_0^\infty dq~
		\frac{1}{q^2 +  \kappa^2  + \frac{|\omega_n|}{\alpha q}}
\nonumber\\&
	\sim
		J\nu \left( \frac{\kappa_0}{k_F} \right)^2
		\frac{k_F}{\kappa + (\omega_n/\alpha)^{1/3}}
	\label{eq:Vsn_2D}
	\end{align}
where $\nu$ is the 2D density of states.\footnote{Eq.~\eqref{eq:Vsn_2D} is not a truncated series expansion but rather an approximate form with the correct limiting behavior at small $n$ and large $n$.}
$\calV^s_n$ diverges as $\omega^{1/3}$, and is cut off for $\omega_n<\alpha\kappa^3$, at a value of the order of $1/\kappa$.
For a 3D spherical Fermi surface, substituting Eq.~\eqref{e:eta_3d} into \eqref{e18} gives
	\begin{align}
 	\calV^s _n
	&=\frac{mJ {\kappa_0}^2}{4\pi^2 k_F}
		\int_0^{2k_F} dq~
		\frac{q}{q^2 + \kappa^2 + \frac{|\omega_n|}{\alpha q}}
\nonumber\\&
	\approx
		\frac{J\nu}{2} \left( \frac{\kappa_0}{k_F} \right)^2
		\ln \frac{2k_F}{\kappa + (\omega_n/\alpha)^{1/3}}
	\label{eq:Vsn_3D}
	\end{align}
with logarithmic accuracy for small, positive $\omega_n$.  Here $\nu$ is the 3D density of states.  The integrand has a divergence cut off by $\omega_n$ and $\kappa^2$. 
$\calV^s_n$ diverges as $\ln 1/\omega_n$, and is cut off for $\omega_n<\alpha\kappa^3$, at a value of the order of $\ln 1/\kappa$.

\subsection{Strong-coupling equations  near FM-QCP}
\begin{figure}\centering
	\includegraphics{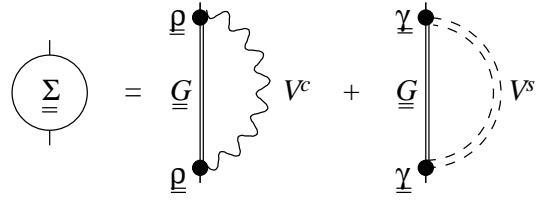}
	\caption{\label{f:eliashberg_diagram} Diagrams for the matrix self-energy.}
	\end{figure}

We now investigate the effect on superconductivity.  
The self-consistent `strong-coupling' equations for the  $4\times 4$ matrix Matsubara Green functions $\bcalG$ and self-energies $\boldsymbol{\Sigma}$ are
	\begin{align}
	\boldsymbol{\Sigma}_{m\kk}
	&=
		T\sum_n \int_\pp
		\left[	
		\calV^c_\stk{m-n}{\kk-\pp}	
		\brho_3
		\bcalG_\stk{n}{\pp}	
		\brho_3
	+
		\calV^s_\stk{m-n}{\kk-\pp}	
		\bgamma_j
		\bcalG_\stk{n}{\pp}	
		\bgamma_j
		\right]
		\\
	\bcalG_{n\kk}
	&=\left(	
			i\vare_n \mathbf{1}	- \xi_\kk \brho_3  
		- \boldsymbol{\Sigma}_{n\kk}
		\right)^{-1}
	\end{align}
where $\brho$ and $\bgamma$ are suitable generalizations of Pauli matrices.\cite{Maki69} 
These equations take into account both the scattering and pairing effects of the phonon-mediated interaction $\calV^c$ and the spin-fluctuation-mediated-interaction $\calV^s$ (although they neglect vertex corrections).
The self-energy equation is represented diagrammatically in Fig.~\ref{f:eliashberg_diagram}.
For isotropic superconductivity with interactions in charge and spin channels the strong-coupling equations reduce to the following frequency-dependent, momentum-independent form:
	\begin{align}
	\label{e42}
	\tvare_m
	&=\vare_m
	+	T\sum_n		\calV^+ _{m-n}		\mathcal{G}_n
	\\
	\label{e43}
	\tDelta_m
	&=	~~~ ~~~		
		T\sum_n		\calV^- _{m-n}		\mathcal{F}_n
	\\
	\label{e44}
	\mathcal{G}_n 
	&=\pi\nu \frac{\tvare_n}{
		\sqrt{ \tvare_n{}^2 + \tDelta_n{}^2 }}
	\\
	\label{e45}
	\mathcal{F}_n 
	&=\pi\nu \frac{\tDelta_n}{
		\sqrt{ \tvare_n{}^2 + \tDelta_n{}^2 }}
	\end{align}
where $\calV^\pm_n = \calV^c_n \pm \calV^s_n$, and $\mathcal{G}_n$, $\mathcal{F}_n$, $\tvare_n$, $\tDelta_n$ are the momentum-integrated ordinary and anomalous Green functions and self-energies.

\subsection{Abrikosov-Gor'kov equations near FM-QCP}
In the vicinity of the QPT, where $\kappa$ is small, $\calV^s_n$ is dominated by the divergence at small $\omega_n$.  Now, $\calV^s_n$ is sampled at discrete bosonic Matsubara frequencies $\omega_n=2\pi T(n+1/2)$.  If the first nonzero Matsubara frequency, $\omega_1 = 2\pi T$, is much larger than the frequency cutoff $\alpha\kappa^3$, we can discard all nonzero Matsubara frequencies.  This corresponds to replacing the dynamic interaction $\calV^s _n$ by a static one,
$ 	\calV^s _n	\approx		\calV^s_0~  		\delta_n.
$.
The spin fluctuations are now characterized by a single number, $\calV^s_0$.
From Eqs.~\eqref{eq:Vsn_2D} and \eqref{eq:Vsn_3D},
	\begin{align}
 	\calV^s_0
	&\approx
		J\nu
		\left( \frac{\kappa_0}{k_F} \right)
		\left( \frac{\kappa_0}{\kappa} \right)
		&\text{(2D model)},
	\label{eq:Vs0_2D}
\\
 	\calV^s_0
	&\approx
		\frac{J\nu}{2} \left( \frac{\kappa_0}{k_F} \right)^2
		\ln \frac{2k_F}{\kappa} 
		&\text{(3D model)}.
	\label{eq:Vs0_3D}
	\end{align}
Let us also replace the dynamic, phonon-mediated interaction by a BCS-type interaction ($\calV^c _n	=	V^\text{BCS}$), with a frequency cutoff $\omega_D$ of the order of the Debye frequency.  Then Eqs.~\eqref{e42} and \eqref{e43} simplify to
	\begin{align}
	\label{e50}
	\tvare_n
	&=\vare_n	
	+	T	\calV^s_0		\mathcal{G}_n
	\\
	\label{e51}
	\tDelta_n
	&=
		T\sum_m^{|\omega_m|<\omega_D}	V^\text{BCS}		\mathcal{F}_m
	-	T \calV^s_0		\mathcal{F}_n.
	\end{align}

Compare these with the case of static charge and spin disorder\cite{Abrikosov61}, that is, magnetic impurities in a superconductor.  That system can be described by the BCS model with additional random delta-correlated potentials $U^c(\xx)$ and $U^s_i(\xx)$ in the charge and spin channels, whose variances are $W^c$ and $W^s$ in the sense that 
$\left< U^c(\xx) U^c(\xx') \right> = W^c \delta(\xx-\xx')$ and 
$\left< U^s_i(\xx) U^s_j(\xx') \right> = W^s \delta(\xx-\xx') \delta_{ij}$ :
	\begin{align}
	S^\text{AG} [\psi,\bar\psi,U^c,U^s]
	&=T\sum_{n\alpha} \int_\kk (i\vare_n - \xi_\kk) 
		\bar\psi_{n\kk\alpha} \psi_{n\kk\alpha}
						\nonumber\\{}
		+\tfrac{1}{2} V^\text{BCS} \int_{\tau\xx} \rho_{\tau\xx}\rho_{\tau\xx}
&
	-	\int_{\tau\xx} U^c(\xx) 		\rho_{\tau\xx}
	-	\int_{\tau\xx}\sum_i U_i^s(\xx) 		S_{\tau\xx i}
	\end{align}
Let $w^{c,s} = 2\pi\nu W^{c,s} = 1/\tau_{c,s}$, where $\tau_{c,s}$ are the ordinary and spin-flip scattering times, and $\nu$ is the density of states at the Fermi energy.  Define $w^\pm = w^c \pm w^s$.  One then obtains the Abrikosov-Gor'kov equations 
	\begin{align}
	\label{e53}
	\tvare_n
	&=\vare_n + 
		\frac{w^+}{2} 
		\mathcal{G}_n
	\\
	\label{e54}
	\tDelta_n
	&=	
		T\sum_m^{|\omega_m|<\omega_D}	  V^\text{BCS}
		\mathcal{F}_m
	+
		\frac{w^-}{2} 
		\mathcal{F}_n
	\end{align}
Comparing Eqs.~\eqref{e50} and \eqref{e51} with \eqref{e53} and \eqref{e54} shows that the quasi-static spin-spin interaction $\calV^s$ can effectively be replaced by static spin disorder whose strength $w^s = T\calV^s _0$ is \emph{temperature-dependent}.  The factor of $T$ arises from the Matsubara frequency sum in the Eliashberg equations, and is necessary on dimensional grounds.
	\begin{figure} 
	\includegraphics[scale=0.5]{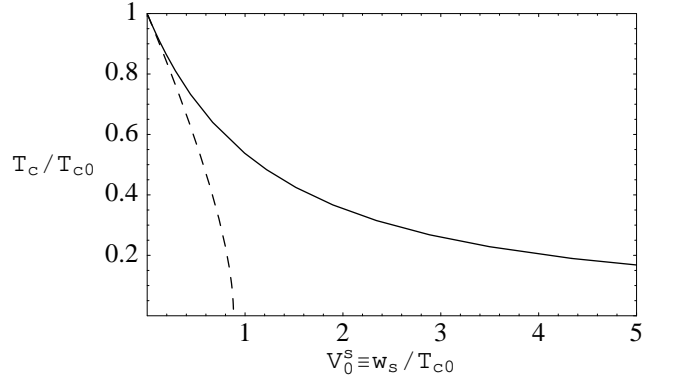}
	\caption{\label{f:abrikosovgorkov} Suppression of $T_c$ by static magnetic disorder of strength $w^s$ (dashed line) and by quasi-static spin fluctuations of strength $V^s_0$ (solid line).}
	\end{figure}

From Abrikosov-Gor'kov theory it is known that $T_c/T_{c0}$ depends only on $w^s$ according to the following \emph{implicit} equation\cite{Abrikosov61}:
	\begin{align}
	\ln\frac{T_c}{T_{c0}}
	&=\psi\left(\frac{1}{2}												\right)
	-	\psi\left(\frac{1}{2} + \frac{w^s}{2\pi T_c} \right)
	\label{e68}
	\end{align}
where $T_{c0}$ is the critical temperature of the clean superconductor, and $\psi(z)$ is the digamma function.
In this scenario, $T_c$ falls to zero when $w^s = \frac{\Delta_{00}}{2} = \frac{\pi}{2e^\gamma} T_{c0}$ (Fig.~\ref{f:abrikosovgorkov}, dashed line).  A \emph{finite} concentration of magnetic impurities is sufficient to destroy superconductivity.  
However, the extra factor of $T$ in the spin-fluctuation scenario indicates that at lower temperatures there are fewer excitations and the depairing effect is less pronounced.  Substituting in Eq.~\eqref{e68}, we find
	\begin{align}
	\ln\frac{T_c}{T_{c0}}
	&=\psi\left(\frac{1}{2}												\right)
	-	\psi\left(\frac{1}{2} + \frac{\calV^s_0}{2\pi} \right)
	\end{align}
This is an \emph{explicit} equation for $T_c$ which is even simpler than the AG result (Fig.~\ref{f:abrikosovgorkov}, solid line)!  The limiting forms are
	\begin{align}
	\frac{T_c}{T_{c0}}
	&=1-\frac{\pi}{4} \calV^s_0 
	\qquad\text{for $\calV^s_0 \rightarrow 0$},
		\\
	\frac{T_c}{T_{c0}}
	&=\frac{\pi}{2e^\gamma \calV^s_0}
	\qquad\text{for $\calV^s_0 \rightarrow \infty$} .
	\label{eq:Tc}
	\end{align}
Superconductivity is only destroyed when the strength of the spin fluctuations becomes infinite.  Inserting Eq.~\eqref{eq:Vs0_2D} or \eqref{eq:Vs0_3D} into \eqref{eq:Tc} gives
	\begin{align}
	\frac{T_c}{T_{c0}}
	&\approx
		\frac{\pi}{2e^\gamma}~
		\frac{1}{J\nu}
		\left( \frac{k_F}{\kappa_0} \right)
		\left( \frac{\kappa}{\kappa_0} \right)
		&\text{(2D)},
	\label{e:Tc_2D}
\\
	\frac{T_c}{T_{c0}}
	&\approx
		\frac{\pi}{2e^\gamma}~
		\frac{2}{J\nu} \left( \frac{k_F}{\kappa_0} \right)^2
		\left(	\ln \frac{2k_F}{\kappa} 	\right)^{-1}
		&\text{(3D)}.
	\label{e:Tc_3D}
	\end{align}

In the immediate vicinity of the QCP, the bare inverse correlation length $\kappa_0$ is a constant of the order of $k_F$, whereas the true inverse correlation length $\kappa$ goes to zero according to some power law, $\kappa \sim |J-J_c|^{\nu}$, as $J$ approaches $J_c$.  In Eqs.~\eqref{e:Tc_2D} and \eqref{e:Tc_3D}, the dimensionless quantities $J\nu$ and $k_F/\kappa_0$ are constants of order unity, whereas the ratio $\kappa_0/\kappa$ (which is the square root of the Stoner enhancement factor) becomes infinite.  Hence, $T_c(J) \sim \frac{1}{\ln |J-J_c|}$ in 3D, and $|J-J_c|^\nu$ in 2D.  Note that the fluctuations are more severe in 2D than in 3D, and $T_c$ is more strongly suppressed.  The quantum critical exponent $\nu$ affects the form of the 2D result but only the prefactor of the 3D result.

It is likely that $(J-J_c)$ is proportional to $(p-p_c)$, where $p$ is an experimental control parameter such as pressure.  Substituting $p$ for $J$, we obtain predictions that can be compared with experiment.  
	
In the limit $\kappa\rightarrow 0$, $1/\ln (1/\kappa)$ and $\kappa$ are always larger than $\alpha\kappa^3$.   Hence our initial assumption $2\pi T \gg \alpha\kappa^3$ is valid in the vicinity of the superconducting transition line.  This justifies the working above \emph{a posteriori}.

Farther away from the QCP, the other factors in Eqs.~\eqref{e:Tc_2D} and \eqref{e:Tc_3D} may come into play.  It is likely that changing the experimental parameter $p$ has a significant effect not only on $J$, but also on the degree of nesting of the Fermi surface, and hence on $\kappa_0$.  Then $T_c$ will deviate from a pure logarithm or power law.

\subsection{Phase diagram}
In the presence of an `externally applied exchange field', such as would arise from artificially polarized impurity spins, it is possible for superconducting order to exist.  In fact, Larkin and Ovchinnikov \cite{Larkin65} show that the exchange splitting $h$ has no effect on $\Delta$ until it reaches about $\Delta/2$.  (The case of an externally applied magnetic field is more complicated due to orbital effects.)  

The case for coexistence of superconductivity and \emph{spontaneous}  magnetization is a more difficult one.  In a fully self-consistent mean-field theory, the superconducting order parameter $\Delta$ and the magnetic order parameter $M$ attempt to suppress each other.  For a single-band model in which the same electrons take part in Cooper pairing and in `Stoner' itinerant ferromagnetism, the competition is so intense that the coexistence region in Fig.~\ref{f21} is eliminated, leaving a first-order FM/SC phase boundary, terminating in a first-order QCP, and the analysis in this paper is not relevant.  It may be possible, however, for FM and SC to coexist in multi-band systems.  \emph{In the following discussion, we presuppose coexistence of FM and SC at the mean-field level.}

First let us assume that the magnetic phase boundary is vertical.
 On the \emph{left} of this boundary, $M=0$, so $T_c$ should be constant (dashed line in Fig.~\ref{f21}).  The magnetic fluctuations should suppress $T_c$ to zero as described above (solid curve).  On the \emph{right} of the magnetic phase boundary, magnetic order is present, and $T_c$ should be suppressed at the mean-field level.  \emph{Suppose that the mean-field values of $T_c$ are given by the dashed curve in Fig.~\ref{f21}}.  Now, since the equation for $T_c$ is linear, the pair-breaking effect of FM order and FM fluctuations combine additively in the argument of the digamma function.  Since the fluctuations are present on both sides of the transition, we expect that $T_c(J)$ is suppressed to zero as $J\rightarrow J_c$ from \emph{either} direction (solid curve):
	\begin{figure} \centering
	\includegraphics{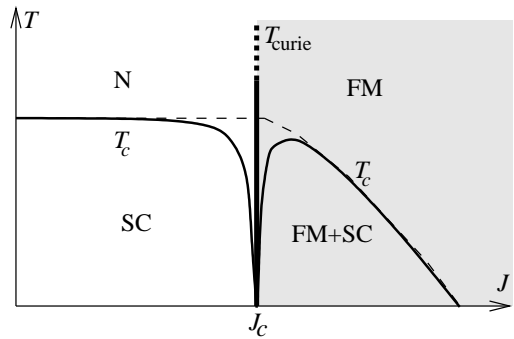} 
	\caption{\label{f21} Corrections to mean-field phase diagram due to fluctuations, assuming vertical FM phase boundary.  `N' represents the normal, non-magnetic metallic phase.}
	\end{figure}

\section{Effect of SDW fluctuations on $T_c$}

\subsection{$\eta_\qq$ near an SDW-QCP}
In order to treat a system close to a SDW instability, we must abandon the assumption of a spherical Fermi surface.  A SDW system typically has a Fermi surface which is almost `nested', meaning that there are portions of the Fermi surface which almost overlap when shifted relative to each other by a nesting vector $\QQ$.  The spin fluctuations with wavevectors close to $\QQ$ are the low-energy quantum critical fluctuations, and it is they that play an important role in suppressing superconductivity.

In order to proceed we need to choose a definite model for the Fermi surface.  Let us begin with a 2D model.
Suppose the Fermi surface is a circle of radius $k_F$.  Shifting this by $2k_F$ in any direction gives a circle tangent to the original circle, suggesting imperfect nesting.  Now the anisotropy of the crystal comes into play.  Since the Fermi surface typically occupies a sizeable fraction of the Brillouin zone, we must consider nesting between periodic images.  In $\kk$-space, shifting the Fermi surface by $\QQ$ makes it tangent to the original Fermi surface at \emph{two} points rather than one.  

\begin{figure} \centering
	\includegraphics[width=90mm]{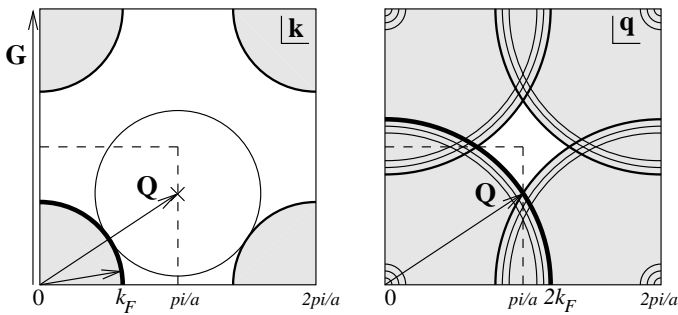}
	\caption{\label{f33} (Left) Circular Fermi surface for $0 \leq k_x, k_y \leq \frac{2\pi}{a}$.  Solid gridlines indicate $k$-space unit cell boundaries;  dashed lines indicate boundary of 1st Brillouin zone.  Empty circle is image of Fermi surface under translation by $\QQ$. (Right) Contours of $\eta_\qq$ (schematic).}
	\end{figure}

It is useful to perform the analysis in $\qq$-space.  The $1/q$ singularity in $\eta_\qq$ is still present, but there is a more relevant $1/\sqrt{2k_F-q}$ singularity on a circle of radius $2k_F$.  This circle and its periodic images intersect at the point $\QQ$ (and three other points related to $\QQ$ by symmetry), and there $\eta_\qq$ is enhanced by a factor of 2.  This construction, illustrated in Fig.~\ref{f33}, determines the optimal nesting vector\cite{Littlewood93}.  In summary, the \emph{radius} of the Fermi surface determines the \emph{magnitude(s)} of the nesting vectors, and the lattice anisotropy determines their \emph{direction(s)}.

	\begin{figure} \centering
	\includegraphics[width=90mm]{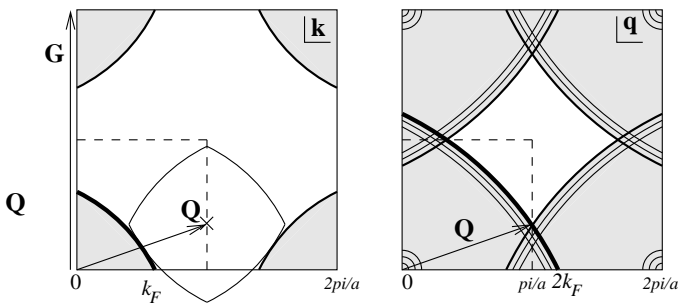}
	\caption{\label{f34} Flattened Fermi surface.  The line singularities in $\eta_\qq$ have a larger magnitude than before.  The singularity at $\qq=\0$ also changes shape, but this is not shown, and is not relevant to the discussion.}
	\end{figure}

Now suppose the Fermi surface undergoes `pincushion' distortion, such that the circular arcs that make up the Fermi surface become flatter (Fig.~\ref{f34}).  The optimal vector $\QQ$ may change; more importantly, the `nesting' singularities in $\eta_\qq$ become larger.  Assuming that the Fermi velocity $v_F$ and the density of states $\nu$ are not significantly altered during the distortion,
	\begin{align}
 	\eta_\qq	
&	\approx
		\frac{m}{2\pi {k_F}^2}~
		\frac{\Theta(1 - \frac{q}{2k_C}) }
			{\frac{q}{2k_C} \sqrt{1 -  \big( \frac{q}{2k_C} \big)   ^2}}
	\approx
		\frac{m \Theta(-q_\perp)}{2\pi {k_F}^2}		
		\sqrt{ \frac{k_C}{q_\perp} }
	\end{align}
where $q_\perp$ is the perpendicular distance from $\qq$ to the Fermi surface in $\qq$-space (towards the center of the arc).  The magnitude of the singularity increases with the square root of the radius of the curvature.  Remember that this is enhanced by a factor of 2 where the singularities intersect.
When $k_C=\infty$ the Fermi surface becomes a square and $\eta_\qq \propto \delta(q_{\perp})$.

In the 3D analogue of this model, the Fermi surface consists of eight, approximately triangular, portions of spherical surfaces.
As the radius of curvature of each portion is increased towards infinity, the Fermi surface morphs from a sphere to an octahedron:
	\begin{align}
 	\eta_\qq	
&\approx
		\frac{m}{4\pi {k_F}^3}~
		\frac{2\pi {k_C}^2 \Theta(1 - \frac{q}{2k_C}) }{q}
	\approx
		\frac{m \Theta(q_\perp)}{4 {k_F}^3}		
		k_C 
		.
	\end{align}
The discontinuity in $\eta_\qq$ is proportional to the radius of curvature.  

\subsection{Spin-spin interaction $\calV^s_{n\qq}$ near an SDW-QCP}
We also need an approximation for the spin-spin interaction $\calV^s_{n\qq}$. 
First consider the bare Green function $\chi^0_{n\qq}$, given by
	\begin{align}
	\chi^0_{n\qq}
	&\propto
		\int_\kk 
		\frac{f(\xi_\kk) - f(\xi_{\kk+\qq})}{i\omega_n - \xi_{\kk+\qq} + \xi_\kk }
	.
	\end{align}
This integral is somewhat similar to the expression for $\eta_\qq$, Eq.~\eqref{e37}.  It is large when $\qq$ is close to a nesting vector of the Fermi surface.  In this case, the definition of `nesting' is more stringent:  portions of the Fermi surface overlap when shifted by the nesting vector $\QQ$, and the `empty' side of one portion matches up with the `filled' side of the other.  Because of this,
$\chi^0_{n\qq}$ does not have a peak near $\qq=\0$, unlike $\eta_\qq$.  We shall assume that both $\chi^0_{n\qq}$ and $\eta_\qq$ are peaked at the same wavevector $\QQ$.  
\emph{This is a reasonable assumption if the SDW and SC are caused by nesting of the same Fermi surface.}

For small $w$ and for $\qq$ close to $\QQ$, $\chi^0_{n\qq}$ can be approximated by a Hertzian of width $\kappa_0$.  In this model, it is expected that $\chi^0 _{0\qq}$ and $\kappa_0$ will both depend upon the curvature parameter $k_C$.  In particular, when $k_C$ is large, the system is well nested, and $\kappa_0$ is expected to be small.
	\begin{align}
 	\chi^0 _{n\qq}
	&=	\frac{\chi^0_{0\QQ} {\kappa_0}^2}
		{{\kappa_0}^2 + |\qq-\QQ|^2 + \frac{|\omega_n|}{\alpha}}.
	\end{align}
Then, in $V^s_{n\qq}$,  $\kappa_0$ is replaced by the `renormalized' inverse spin-fluctuation correlation length $\kappa$, which is smaller due to Stoner enhancement:
	\begin{align}
 	\calV^s _{n\qq}
	&=	\frac{J {\kappa_0}^2}{\kappa^2 + |\qq-\QQ|^2 + \frac{|\omega_n|}{\alpha}}
	\end{align}
where $\kappa = {\kappa_0} \left( 1 - J\chi^0_{0\QQ}  \right)^\nu$.  Ultimately, the FLEX calculation requires only the $n=0$ component, which is a Lorentzian in $\qq$-space peaked at $\qq=\QQ$:
	\begin{align}
 	\calV^s_{0\qq}
	&=	\frac{J {\kappa_0}^2}{\kappa^2 + |\qq-\QQ|^2}.
	\end{align}

	\begin{figure} \centering
	\includegraphics[width=90mm]{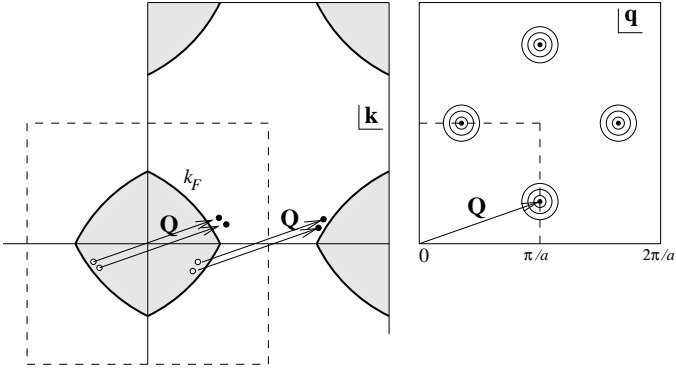}
	\caption{(Left) Electron-hole excitations near the nesting vector which give the main contribution to $\chi$.  (Right) Schematic contour plot of $\chi_{0\qq}$ (or $V^s_{0\qq}$).}
	\end{figure}

\subsection{Effective interaction $\calV^s _n$ near SDW-QCP;  suppression of $T_c$}
The momentum-averaged interaction is given, as before, by Eq.~\eqref{e18}.  Since $\eta_\qq$ has a weaker divergence for the SDW case than for the FM case, $\calV^s_{0}$ also has a weaker divergence, as we shall see below.  First consider the 2D model.  To do the integral over $\qq$, we use coordinates $q_\parallel$ and $q_\perp$ oriented tangentially and perpendicularly to the Fermi surface.  There is a factor of 2 arising from the overlap of two singularities.
	\begin{align}
 	\calV^s_n
	&=\int \frac{d^2q}{(2\pi)^2}~	\eta_\qq~ \calV^s _{n\qq}
	\nonumber\\
	&=\frac{1}{(2\pi)^2}
		\int dq_\perp \int dq_\parallel 
		\frac{m}{2\pi {k_F}^2}~
		\Theta(q_\perp)
		\sqrt{ \frac{k_C}{q_\perp} }
\times\nonumber\\{}&~~~~~~~~~~~~~~~~~~~~~~~~~\times
		\frac{J {\kappa_0}^2}
			{\kappa^2 + {q_\parallel}^2 + {q_\perp}^2 + \frac{|\omega_n|}{\alpha}}
		\times 2
	\nonumber\\
	&\approx
		\frac{mJ {\kappa_0}^2 \sqrt{k_C}}{2\pi^3 {k_F}^2}
		\int_0^\infty \frac{dq_\perp}{\sqrt{q_\perp}}~
		\int_{-\infty}^\infty 
		\frac{dq_\parallel}
			{{q_\parallel}^2 + {q_\perp}^2 + \kappa^2 +  \frac{|\omega_n|}{\alpha}}
	\nonumber\\
	&\approx
		\frac{J\nu}{\pi}~ \frac{{\kappa_0}^2}{{k_F}^2} \sqrt{k_C}
		\int_0^\infty \frac{dq_\perp}{\sqrt{q_\perp}}~
		\frac{\pi}
			{\sqrt{{q_\perp}^2 +\kappa^2 +  \frac{|\omega_n|}{\alpha}}}
	\nonumber\\
	&\approx
		\frac{J\nu}{\pi}~ \frac{{\kappa_0}^2}{{k_F}^2} \sqrt{k_C}
		\frac{8 \Gamma(5/4)^2}{
				\sqrt{\pi} \left(\kappa^2 +  \frac{|\omega_n|}{\alpha}  \right)^{1/4}}
		.
	\end{align}
At zero frequency,
	\begin{align}
 	\calV^s_0
	&\sim
		J\nu~ \frac{{\kappa_0}^2}{{k_F}^2} 		\sqrt{\frac{k_C}{\kappa}}.
	\label{e83}
	\end{align}
The strength of the effective interaction increases when the Fermi surface becomes flatter or when the coupling $J$ is tuned towards $J_c$.  In general, $\kappa$ goes to zero at the QCP whereas $k_C$ is large but finite.  Therefore, near the QCP, $ 	\calV^s_0
	\sim 1/\sqrt{\kappa}$,
and by the arguments of the previous sections, the superconducting $T_c$ is suppressed as
	\begin{align}
 	T_c
	&\sim \sqrt{\kappa} 
	\sim (J-J_c)^{\nu/2}.
	\end{align}
The quantity $J\nu$ is of order unity, but $\kappa_0/\kappa_F$ is likely to be small, as explained earlier.  Hence the suppression of $T_c$ is much weaker for the SDW case than for the FM case.

For the 3D model, an estimate gives
	\begin{align}
 	\calV^s_n
	&=\int \frac{d^3q}{(2\pi)^3}~	\eta_\qq~ \calV^s _{n\qq}
	\nonumber\\
	&=\frac{1}{(2\pi)^3}
		\int_{0}^{k_F} dq_\perp \int dq_\parallel~ 2\pi q_\parallel~
		\frac{m}{4 {k_F}^3}~
		\Theta(q_\perp)
		k_C
\times\nonumber\\{}&~~~~~~~~~~~~~~~~~~~~~~~~~~\times
		\frac{J {\kappa_0}^2}
			{\kappa^2 + {q_\parallel}^2 + {q_\perp}^2 + \frac{|\omega_n|}{\alpha}}
		\times 4
	\nonumber\\
	&\sim
		\frac{mJ{\kappa_0}^2  k_C}{{k_F}^3}
		\int_{0}^{k_F} dq_\perp~
		\int_{-k_F}^{k_F}
		\frac{dq_\parallel~ q_\parallel}
			{{q_\parallel}^2 + {q_\perp}^2 + \kappa^2 +  \frac{|\omega_n|}{\alpha}}
	\nonumber\\
	&\sim
		\frac{J\nu{\kappa_0}^2  k_C}{{k_F}^4}
		\int_{0}^{k_F} dq_\perp~
		\ln	\frac{k_F}{
				\sqrt{{q_\perp}^2 +\kappa^2 +  \frac{|\omega_n|}{\alpha}}
			}
	\nonumber\\
	\therefore\quad
	\calV^s_0
	&\sim
		J\nu \frac{{\kappa_0}^2 }{{k_F}^2}~ \frac{k_C}{k_F}
		\Big(			1	- \frac{\pi}{2}~ \frac{\kappa}{k_F}	\Big)
		.
	\end{align}
The integral is not divergent at small $q_\perp$, so $\calV^s_0$ remains finite, and $T_c$ is not suppressed to zero.  In fact, since $\frac{\kappa_0}{k_F}$ is small, the suppression should be a small effect.  Taking into account the terms of order $\kappa$, we see that 
$\calV^s_0 \sim \text{const} - \kappa \sim \text{const} - |J-J_c|^{\nu} $, so
$T_c \sim \text{const} + |J-J_c|^{\nu}$:  the critical temperature has a cusp near the QCP.

The above arguments apply close to the QCP.  Farther away from the QCP, the variation of $\kappa_0$ may be significant.  In fact, it is likely that the quantum phase transition is caused by an increase in nesting, so $\kappa_0$ decreases as the QCP is approached from far.  
Then the argument implies that $T_c$ should \emph{increase} at first as the QCP is approached.  This can be justified as follows: although Fermi-surface nesting causes an enhancement of $\chi^0$ and $\chi$, and indirectly enhances the $\calV^s$, it also narrows the peak in the interaction in $\qq$-space so that the magnetism competes less effectively for the Fermi surface.  Of course, this result is model-dependent.

\section{Numerical results} 
\begin{figure*}[!ht] \centering
\subfigure[Linear plot of $T_c(\kappa^2)$]{\includegraphics[width=75mm]{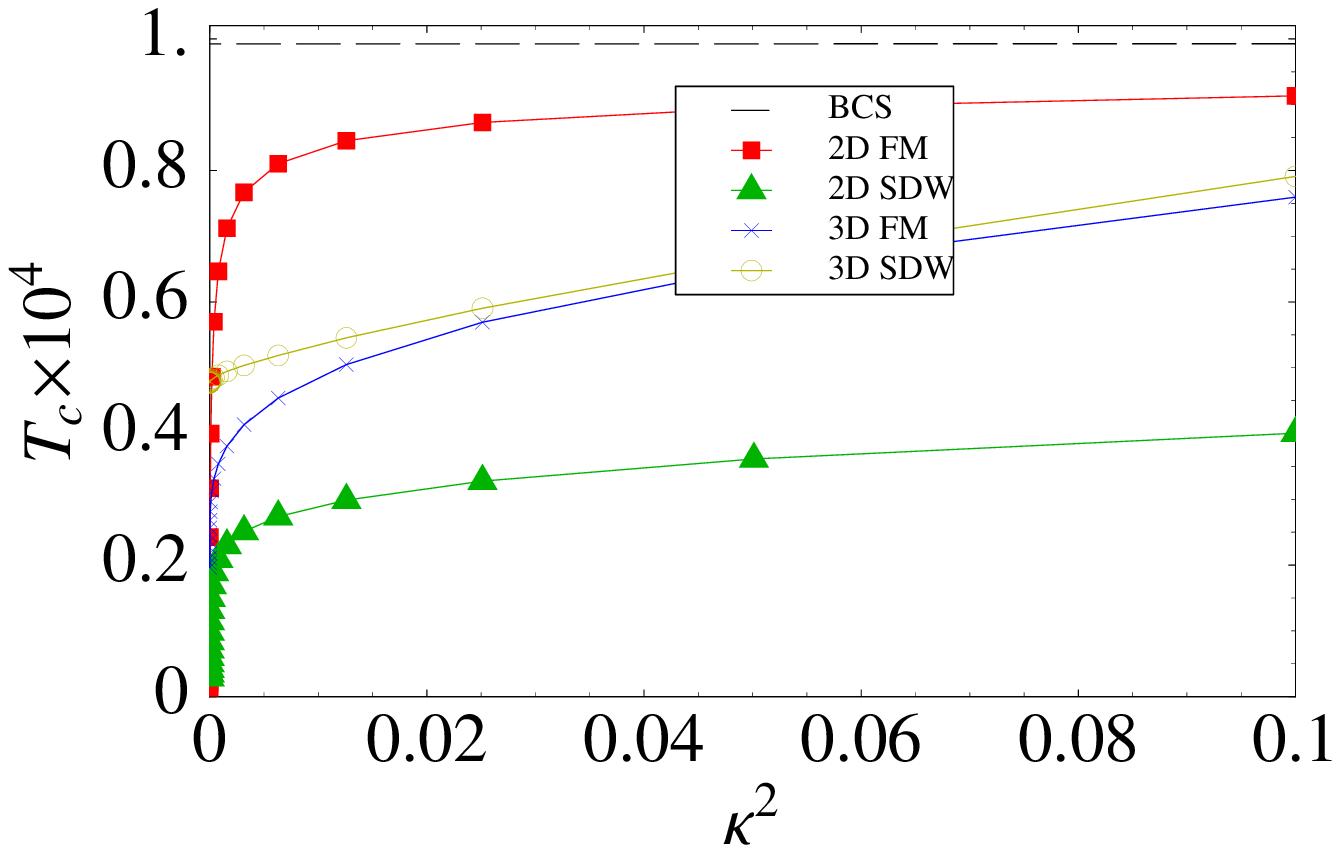}  }
\subfigure[Log-log plot of $T_c(\kappa^2)$]{\includegraphics[width=75mm]{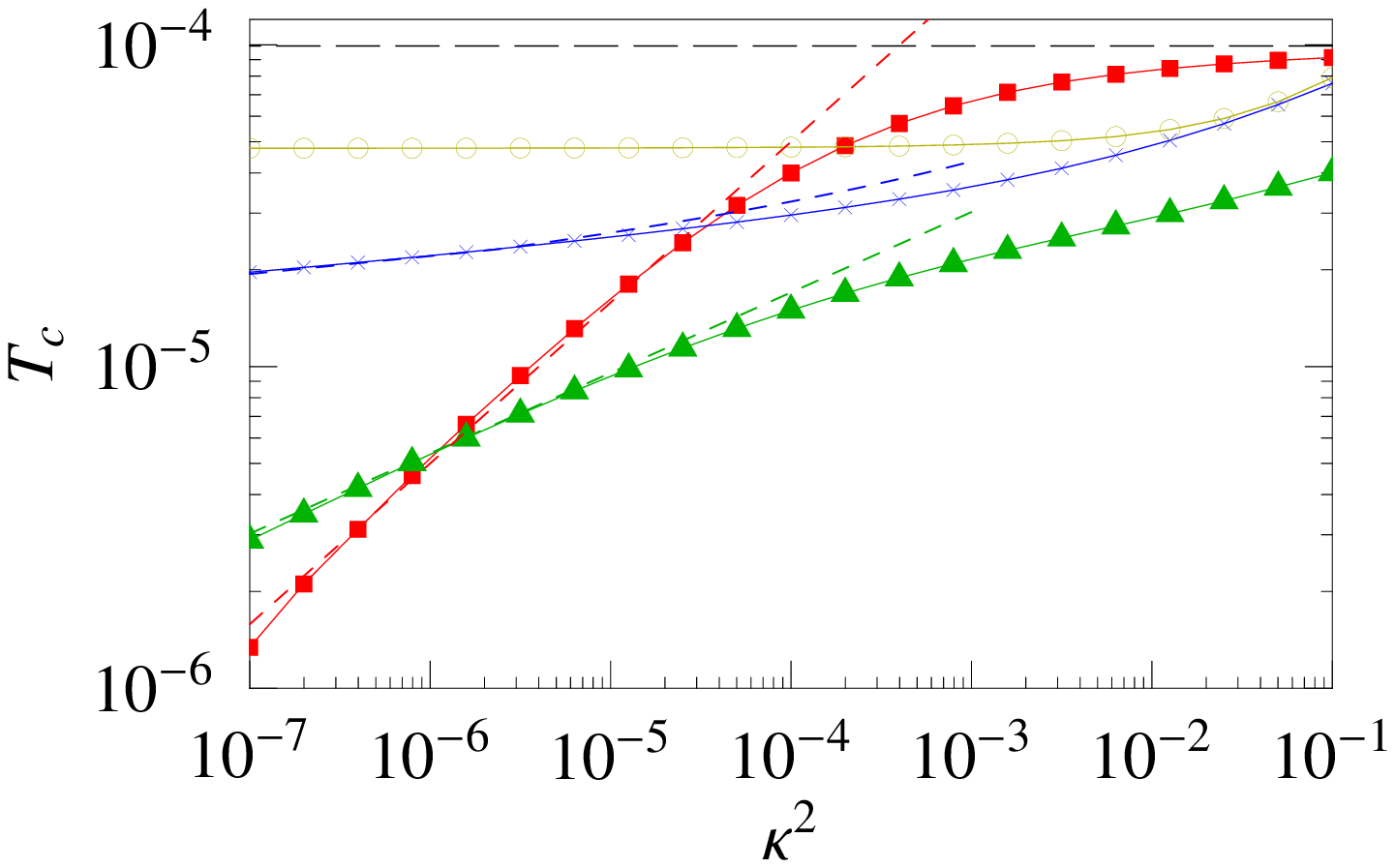} \label{f:loglog}  }
\caption{\label{f:numerics} Numerical results for superconducting $T_c$ as a function of distance from quantum critical point $\kappa^2$ for suitable model parameters, obtained using an imaginary-time Eliashberg approach.}
\end{figure*}

Figure~\ref{f:numerics} shows numerical results obtained by self-consistent solution of Eqs.~\eqref{e42}-\eqref{e45}, using an Einstein phonon model for $\calV^c_n$ and the various approximations derived above for $\calV^s_n$.  From the log-log plot (Fig.~\ref{f:loglog}) it is clear that $T_c$ obeys the expected power laws ($\kappa$ and $\sqrt{\kappa}$) near a 2D FM or AF transition respectively, and that in 3D the suppression of $T_c$ is very weak.

\section{Discussion}
We have made predictions about the analytic behavior of the $T_c$ of an isotropic superconductor near ferromagnetic or spin-density-wave QCPs:
\begin{center}
\begin{tabular}{lll}
2D FM  &~~~& $T_c \sim \kappa           \sim |J-J_c|^\nu$ \\
2D SDW && $T_c \sim \sqrt{\kappa}    \sim |J-J_c|^{\nu/2}$ \\
3D FM  && $T_c \sim 1/\ln(1/\kappa)        \sim 1/\ln (1/|J-J_c|)$ \\
3D SDW && $T_c \sim \text{const} +  |J-J_c|^\nu$ \\
\end{tabular}
\end{center}

$T_c$ is suppressed all the way to zero by ferromagnetic quantum critical fluctuations in 2D and 3D, even before magnetic order sets in.  Spin-density-wave fluctuations have a weaker effect; in 2D they can suppress $T_c$ to zero, but in 3D they merely produce a cusp.  These predictions rely on the assumption of coexistence of magnetism and homogeneous, isotropic superconductivity at the mean-field level.
The results for the 3D FM should be contrasted with the prediction in the $p$-wave case of a finite $T_c$ at the QCP.\cite{Wang01}

These results are based on the theory of Abrikosov and Gor'kov, which has undergone several refinements since its publication in 1961.  Calculations including higher-order scatterings from a classical spin\cite{Yu65,Shiba68,Rusinov69} and non-perturbative studies \cite{Lamacraft01} suggest that magnetic impurities lead to states within the superconducting gap; however, the predicted $T_c$ (which is all that matters to our conclusions) is not qualitatively different from that of Abrikosov and Gor'kov.
Taking the quantum nature of the spins into account\cite{muellerhartmann71} suggests that the spins are quenched by the Kondo effect at low temperatures (in the case of antiferromagnetic exchange $J>0$), and thus cease to cause pair-breaking, so that superconductivity is destroyed only at an infinite concentration of magnetic impurities.  However, the present paper deals not with impurity spins but with quantum critical spin fluctuations:
%
in a nearly magnetic metal these are \emph{not} destroyed at low temperatures by any Kondo effect, so there is no reason to expect such a phenomenon in a nearly magnetic superconductor.

We are not aware of any experimental systems where a FM QCP has been seen in a phonon-mediated superconductor.
But, as Saxena et al remark in Ref.\ \cite{Saxena00}, very few itinerant-electron ferromagnets studied to date have been prepared in a sufficiently pure state, or have been `tuned', to be sufficiently close to the border of ferromagnetism, or cooled to sufficiently low temperatures, to provide a definitive check of the predictions of theory.

In Fe \cite{Shimizu01}, signs of superconductivity have been observed on the high-pressure side of a first-order structural phase transition, but it is unclear if this is (i) conventional phonon-mediated superconductivity which is suppressed by ferromagnetism in the bcc phase and revealed in the hcp phase, (ii) unconventional superconductivity mediated by ferromagnetic spin fluctuations, or (iii) by some other type of spin fluctuation present in the hcp phase.  As the transition is first-order, the analysis in this paper is not applicable.

The compound MgCNi$_3$ is a superconductor below at 8K \cite{Lin03}, with a BCS-like ($s$-wave-like) specific heat, and there has been a theoretical prediction that it is unstable to ferromagnetism upon doping with 12\% Na or Li.  Such a ferromagnetic quantum critical point may be a testing ground for this theory.

In ErRh$_4$B$_4$\cite{Fertig77} and HoMo$_6$S$_8$\cite{Lynn81,Ishikawa77} the FM state destroys superconductivity at sufficiently low temperatures, whereas at intermediate temperatures the coexistence gives rise to a compromise oscillatory behavior.  The physics here is dominated by orbital diamagnetism, rather than by paramagnetic spin fluctuations.

The rare-earth nickel borocarbides\cite{Cava94} show coexisting SDW and superconductivity, as well as corresponding nesting features in the Fermi surface.\cite{Dugdale99}  Suppression of $T_c$ using hydrostatic and chemical pressure to manipulate the exchange interaction has been reported\cite{Michor00} but the AFM QCP has not been explored.

In the Bechgaard salt (TMTSF)$_2$PF$_6$, pressure has been used to tune the SDW state through a QCP at $6.4$kbar where a superconducting state is found.  The quasi-2D $\kappa$-BEDT-TTF salts have a high superconducting temperature and are near the boundary of an AF insulating phase.\cite{Akutsu00}  However, the character and mechanism of superconductivity in these organic materials is still debated, and thus the applicability of our model is in doubt.

To conclude, while there are several candidate model systems with which to test these predictions, up to now no sufficiently detailed studies have been performed.

\begin{acknowledgments}
It is a pleasure to thank A.~Nevidomskyy for helpful discussions.  Y.~L.~L. thanks Trinity College, Cambridge for financial support.
\end{acknowledgments}
\bibliography{flex}
\end{document}